# HOW DOES IMAGING IMPACT PATIENT FLOW IN EMERGENCY DEPARTMENTS?

Vishnunarayan Girishan Prabhu

Kevin Taaffe
Marisa Shehan

Systems Enginnering and Engineering Management
University of North Carolina at Charlotte
245 Cameron Hall
Charlotte, NC 28262, USA

Department of Industrial Engineering
Clemson University
100 Freeman Hall
Clemson, SC 29631, USA

Ronald Pirrallo
William Jackson
Michael Ramsay
Jessica Hobbs

Department of Emergency Medicine
Prisma Health - Upstate
701 Grove Road
Greenville, SC 29605, USA

## ABSTRACT

Emergency Department (ED) overcrowding continues to be a public health issue as well as a patient safety issue. The underlying factors leading to ED crowding are numerous, varied, and complex. Although lack of in-hospital beds is frequently attributed as the primary reason for crowding, ED's dependencies on other ancillary resources, including imaging, consults, and labs, also contribute to crowding. Using retrospective data associated with imaging, including delays, processing time, and the number of image orders, from a large tier 1 trauma center, we developed a discrete event simulation model to identify the impact of the imaging delays and bundling image orders on patient time in the ED. Results from sensitivity analysis show that reducing the delays associated with imaging and bundling as few as 10% of imaging orders for certain patients can significantly (p-value < 0.05) reduce the time a patient spends in the ED.

## 1  INTRODUCTION

Emergency departments (EDs) act as one of the primary patient care access points for millions of people seeking medical care. The ever-increasing volume of patient arrivals and varying severity among cases makes ED one of the most complex healthcare settings and prone to crowding (Center for Disease Control and Prevention 2015). Crowding is a well-recognized public health and patient safety issue that has been explored over the last few decades, which occurs when the patient demand for emergency care exceeds the resources available in the ED to provide care in an acceptable time period (American College of Emergency Physicians 2019; Hoot and Aronsky 2008; Morley et al. 2018). ED crowding has a negative impact on patients, providers, and health systems where it leads to reduced quality of care, poor patient outcomes, increased medical errors, higher patient mortality, increased stress and burnout among providers, and increased healthcare costs (Derlet and Richards 2000; Morley et al. 2018; Trzeciak and Rivers 2003). Despite the public awareness and significant efforts by researchers/government agencies, crowding still plagues EDs across the globe and has risen over the past several years.

Although various reasons contribute to ED crowding, one of the primary reasons leading to ED crowding within the US is the overwhelming increase in patient arrivals to the ED, which has increased by 24%, and the decrease in the number of EDs, which declined by 15% over the last decade (CDC 2010). While typically associated with poor ED design and/or inefficient flow, crowding in the ED is often directly attributed to hospitals running at high census (patient arrivals) or having limited inpatient bed capacity for any number of reasons (Moskop et al. 2009). Prior studies have reported the availability of inpatient beds as a significant factor (but not the only one) contributing to ED crowding (Morley et al. 2018). According to the latest reports, approximately 70% of inpatient hospital admission occurs through the ED (Augustine 2019). Hence a lack of inpatient beds due to various factors, including scheduled admissions like post elective surgery, etc., causes a bottleneck in the ED where the patients who require an in-hospital admission wait on ED beds, making them unavailable to new patients and eventually leading to crowding (George and Evridiki 2015). Apart from these two contributors, additional factors that contribute to crowding include hospital type (i.e., teaching hospitals, trauma centers, and public hospitals have higher levels of ED crowding), lack of access to primary care, inadequate staffing, limited access to psychiatric services, ED bed shortages resulting in a higher length of stay (LOS) and slower discharge rates and ED's dependencies on external departments/resources (Derlet and Richards 2000; Morley et al. 2018).

Numerous simulation/mathematical modeling studies have explored the issues of inadequate staffing, lack of ED beds, and increasing patient arrivals to the ED, including our prior work (Corlu et al. 2020; Goldman et al. 1968; Prabhu et al. 2019; Prabhu et al. 2021). However, to our knowledge, none of the studies have focused on investigating the impact of external resources. We define external resources as any resources not dedicated to ED and shared across the hospital, including imaging, labs, etc. In this research, we focus on the impact of delays resulting from the imaging process and bundling the imaging orders on patient flow in the ED. Although findings from this study are specific to the imaging process and delays associated with a specific hospital, the sensitivity analysis provides an insight into the impact of delays associated with any external resources. In this research, we develop a Discrete Event Simulation (DES) model with both patients and physicians as individual agents/entities to identify the impact of delays resulting from the imaging process and the impact of bundling the imaging orders on patient flow in the ED. The paper is organized as follows: in the next section, we discuss the background and motivation for this study, followed by the data used in the model, model development, and finally, we discuss the results and future directions.

## 2    BACKGROUND

The demand for imaging services by EDs has significantly increased over the years, and according to the Centers for Disease Control and Prevention (CDC) reports, half of all ED visits in the US involved some type of imaging (CDC 2015). Prior studies have observed that the imaging services required increased as the patient severity increased, represented by the Emergency Severity Index (ESI) (Yoon et al. 2003). ESI is a severity score assigned to a patient during the triaging process in the ED, where ESI 1 represents an urgent case that requires immediate care and ESI 5 represents the least urgent case. Imaging has a significant impact on patient flow in the ED and patient LOS, and multiple studies have identified these resources as "bottlenecks" causing delays in the ED (Yoon et al. 2003). However, except for Computed Tomography (CT) scans and Magnetic Resonance Imaging (MRI), which require a lot of time to complete the actual test, understanding the causes of delay related to imaging is complex as there are multiple steps involved (Hurlen et al. 2010). On a high level, an imaging process involves three major steps, a provider placing the order, the radiology team completing the imaging on the patient, and interpreting the results to provide follow-up care. Between these steps, there are two main undesirable intervals (delays): a) time from placing the order to beginning the imaging process and b) time from completing the test until reading the results.

A study investigating the root cause of delays associated with completing an imaging order placed from the ED observed processing of imaging order, transport times, and radiology suite location as the three root causes (Worster et al. 2006). These factors are the major contributors to the wait time before and after the actual imaging process, which is undesirable and has a significant impact on the LOS. From a clinical standpoint, a study compared the overall LOS for patients who received a Point of Care (POC) testing to those who received scans from the radiology department and observed that ED LOS was 120 minutes longer for patients who did not use POC testing (Wilson et al. 2016). Additionally, similar observations were reported by another study that compared POC testing to those performed in the radiology department among children with suspected appendicitis (Elikashvili et al. 2014). Although various researchers have explored similar studies, including comparing patient LOS with POC vs. going to the radiology suite, classifying necessary vs. unnecessary imaging, involving patients in care procedures, etc., very few have tried to identify the impact of wait times (delays) and budling of image orders on time a patient spends in the ED.

Despite being one of the most complex modeling techniques, the capability of simulation models to perform "what-if" analyses without the cost and difficulty associated with a full-scale implementation makes it one of the most effective tools to test the impact of various policies. Simulations models, specifically Discrete Event Simulation (DES), have been utilized to address various issues pertinent to the ED, including resource allocation, patient streaming, ED staffing, the impact of fast-track ED, patient safety, design comparison, etc. (Ahsan et al. 2019; Vanbrabant et al. 2019; Yousefi et al. 2020; Elbeyli and Krishnan 2000; Harper and Shahani 2002). Among these, the first application dates over five decades ago, when a DES model was developed to identify the impact of bed allocation and utilization (Goldman et al. 1968). Further, as mentioned earlier, studies have used DES to identify bottlenecks in the ED, the importance of each resource, and the impact of resources on the performance metrics ( Srinivas et al. 2020; Duguay and Chetouane 2007). Although multiple studies have been conducted, very few studies have explored the issue of imaging delays in the ED. We identified one study that investigated the impact of resources on accessibility for computed tomography (CT) scanning in the ED (Maass et al. 2022). However, this was restricted to one specific type of imaging and focused on what was causing the delay in imaging rather than its impact on the patient time in the ED.

In recent years, the use of Agent-Based Modeling (ABM) has gained a lot of popularity in healthcare operations, including ED. Applications of ABM in ED include evaluating fast-track strategies, physician scheduling, resource allocation, etc. (Adleberg et al. 2017). The capability of ABM to simulate the actions of each independent individual, including physicians, nurses, patients, etc., and their combined actions makes it an effective modeling approach while aiming for high fidelity. In this research study, we developed a DES model representative of our partner ED where physicians and patients are modeled as individual agents to identify the impact of imaging delays on patient time in the ED. Further, we investigate the impact of bundling imaging orders compared to placing separate orders on patient time in the ED.

## 3 METHODS

Our primary aim was to investigate the impact of imaging delays on patient flow in the ED at Prisma Health - Greenville Memorial Hospital (GMH), Greenville, SC. Prisma Health is the largest healthcare provider in South Carolina and serves as a tertiary referral center for the entire upstate region. The flagship GMH academic Department of Emergency Medicine is an Adult Level 1 Trauma Center that provides medical care for over 110,000 patients annually. The partner ED can perform a variety of imaging orders, including X-ray, CT, Ultrasound (US), and MRI. Among these, X-rays and US imaging are the most common and can be performed in the ED. However, for complex MRIs and CT scans, patients are rolled out of the ED to shared resources where ED patients are prioritized for scanning. As some patients require multiple imaging, we investigate how bundling the imaging orders affects the patient time in the ED. To gain a high-level understanding of these delays, we consider specific steps and processes associated with imaging to capture all delays and replicate the partner hospital.

### 3.1 Data and Model Inputs

The data used in this study, including the patient arrivals, ESI level of the patients, imaging orders by ESI, imaging times, delays associated with the imaging process, patient time in the ED, number of beds, physician shifts, and number of interactions between physicians and patients, were gathered from the partner hospital. Additionally, the research team included ED physicians working in the partner hospital to provide guidance, review assumptions, and address any other ED activities included in the model. This study was provided an expedited IRB approval by Clemson University IRB under record IRB2020-101.

Among various input data used in the simulation model, we first discuss the average daily patient arrivals to the partner ED for the time period (July 2019 – September 2019). This period was used based on ED stakeholders' suggestions to avoid variability in patient arrivals observed during the COVID-19 pandemic. Additionally, these months had the most patient arrivals over the last three years. The patient arrivals pattern observed in the partner ED aligns with some of the publicly available datasets where patient arrivals are low during the early hours and slowly start picking up from 7:00 am until 12:00 pm, when they reach the maximum and stay almost the same until 7:00 pm. Further, similar to prior observations, weekdays have higher patient arrivals compared to the weekends, and Mondays have the highest patient arrivals.

The next input data was the patient severity level, and as represented in Table 1, most patients are assigned ESI 3 followed by ESI 2 and 4, and very few patients are assigned ESI 1 and 5. Consistent with the literature, we observed that about 60% of the patients arriving at the partner ED required at least one imaging (Center for Disease Control and Prevention 2015). Further, it can be observed that about 91% of ESI 1 (most severe) patients required at least one imaging compared to only 5.7% of ESI 5 (least urgent) who required imaging. This observation was also in line with literature where imaging requirements are influenced by ESI level, where the likelihood of imaging increased with patient severity (Yoon et al. 2003).

Table 1: Patient severity, arrivals, and imaging requirements.

| Severity | Percent Contribution to Patient Arrivals | Percent requiring at least one image | Imaging orders | | | |
|---|---|---|---|---|---|---|
| | | | 0 | 1 | 2 | 3 |
| ESI 1 | 3.1% | 90.7% | 9.3% | 63.3% | 22.2% | 5.2% |
| ESI 2 | 23.5% | 68.2% | 31.8% | 52.0% | 13.3% | 2.9% |
| ESI 3 | 47.2% | 64.0% | 36.0% | 52.0% | 10.0% | 2.1% |
| ESI 4 | 22.8% | 37.2% | 62.8% | 34.0% | 2.8% | 0.0% |
| ESI 5 | 3.4% | 5.7% | 94.3% | 5.7% | 0.0% | 0.0% |

On investigating the variability among the number of imaging required, we observed that some patients required as many as 6 images. However, further analysis demonstrated that it is not necessarily the number of images but the number of image orders that influences the patient time in the ED, as seen in Table 2. From Table 2 it is evident that irrespective of the number of images required, the time in the ED did not vary significantly (p-value > 0.05) for the same imaging orders. This suggests that even if multiple images are placed under the same order, the patient time in the ED did not vary significantly.

Table 2: Imaging orders, number of images, and patient time in the ED.

| Imaging Orders | Patient Time in ED (mins) for | | | | Average Patient Time in ED (mins) |
|---|---|---|---|---|---|
| | 1 image | 2 images | 3 images | 4+ images | |
| 0 | -- | -- | -- | -- | 155 |
| 1 | 190 | 199 | 203 | 192 | 192 |
| 2 | -- | 266 | 254 | 253 | 258 |
| 3 | -- | -- | 320 | 312 | 318 |

However, as seen in Table 2, the patient time in the ED varied significantly based on image orders (p-value = 0.01). Image orders are defined as separate orders placed by a physician for the same patient at least 20 minutes apart, and this is critical because a single imaging order can include multiple imaging requirements. This decision was made based on expert opinion because ED physicians suggested that after an initial interaction with the patient, the physician could place imaging orders, and after subsequent visits, additional orders could be placed. Further, this approach aligns with the ED operations where a patient is visited multiple times by a physician before discharging or admitting, during which tests and labs are ordered. Table 1 above represents the number of imaging orders required for each ESI level. In our simulation model, the number of patient-physician interactions is based on the ESI level, where an ESI level 1 patient would have 4 interactions, ESI 2&3 would have 3 interactions, and ESI 4&5 would have 2 interactions. These numbers were based on observational studies at the partner ED and feedback from physicians.

The time spent for each interaction was inputted as various probability distributions, and this time was between 15-30% of the time in ED, as represented in Table 3 below, which was based on literature and discussions with physicians (Füchtbauer et al. 2013). In Table 3, TRIA represents the triangular distribution, a type of continuous probability distribution where TRIA (a,b,c) represents a distribution with a lower limit a, upper limit b, and mode c, where a < b and a ≤ c ≤ b. The total time a patient spends in the ED can be largely split into two: "Bed to Disposition" and "Disposition to Discharge/Admit." Bed to Disposition represents the time a patient occupies an ED bed and is provided care by physicians and other medical providers, including imaging, providing medicines, blood draws, etc. Although patients will be waiting in their beds during this period without receiving direct care, all these delays are due to waiting for their test results, medicines, etc. In general, this represents the period a patient first occupies a bed in the ED until the physicians make a disposition decision (admit, discharge or transfer). The second part, "Disposition to Departure," is the period for which a patient occupies the ED bed from the time the physician makes a disposition decision until they are physically moved from the ED (discharged, admitted, or transferred). These represent the logistical delays where a patient can be either waiting until a bed is available in the hospital (admission) or waiting for transportation (discharge or transfer).

Table 3: Patient time with physician for each interaction.

| Severity | Interaction 1 | Interaction 2 | Interaction 3 | Interaction 4 |
|---|---|---|---|---|
| ESI 1 | TRIA (13,14,15) | TRIA (8,9,10) | TRIA (8,9,10) | TRIA (2,3,4) |
| ESI 2 | TRIA (9,10,11) | TRIA (15,16,17) | TRIA (7,8,9) | -- |
| ESI 3 | TRIA (8,9,10) | TRIA (14,15,16) | TRIA (7,8,9) | -- |
| ESI 4 | TRIA (10,11,12) | TRIA (5,6,7) | -- | -- |
| ESI 5 | TRIA (10,11,12) | TRIA (6,7,8) | -- | -- |

Finally, for the imaging process, as briefly discussed in the earlier section, from the time after a physician place an imaging order until the image is read, the time is divided into three sections. The three periods are a) Order to Begin, b) Begin to End, and c) End to Read. Order to Begin is the initial wait time/delay a patient must wait in bed before the imaging process is started. Begin to end represents the actual imaging time, and finally, End to Read represents the time a patient must wait after completing the imaging for the provider to read the results. Here Order to Begin and End to Read time intervals (delays) are undesirable and influenced by various factors.

## 4    SIMULATION MODEL

To develop a simulation model representative of the partner health's ED, we utilized a modeling approach where both patients and physicians are represented as agents with unique attributes in a DES model. Usually, in a traditional DES model of ED, physicians are modeled as resources that a patient (agent) seizes for a specific duration (either constant or stochastic) and later releases, making it available for the next patient. Additionally, most prior simulation studies consider a single patient-physician interaction, whereas, in reality, a patient could have multiple interactions with a physician during a single ED visit. Our modeling approach allowed to replicate the physician activities in the ED in a realistic manner, including starting a shift at a particular time, spending time in their workstation ordering tests, updating a patient record, multiple interactions with a patient, and finally handing off a patient to the next physician when their shift ends. Additionally, this modeling approach enabled to replicate a one-to-many scenario as observed in the ED where multiple patients are provided care by a single physician, but a patient is cared for by a single physician. Although some of these could have been incorporated with a traditional DES modeling, our approach allowed for more flexibility and tracking of each agent. However, it should be noted that in this model, there are no specific inter-agent interactions and collaborative decision-making. Figure 1 below provides an overview of patient flow and physician activities in the ED for a single pod which was developed by conducting observations and meetings with ED physicians.

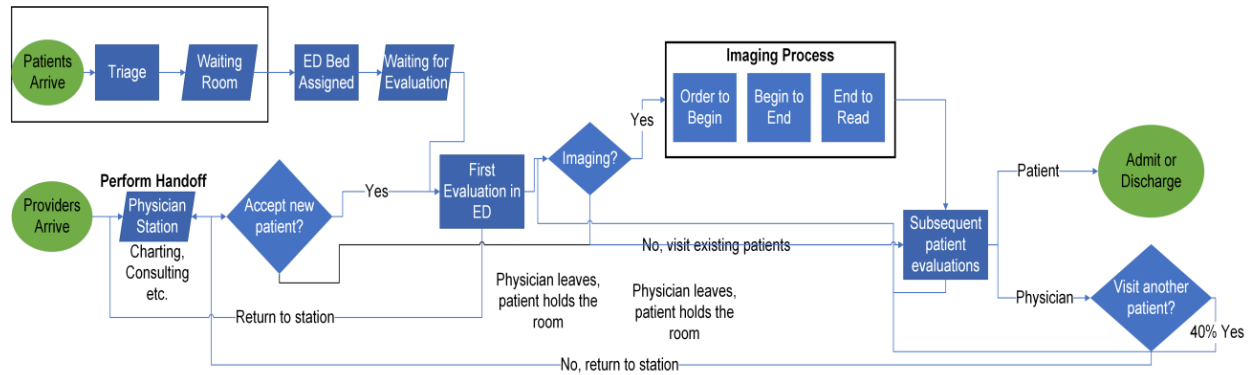

Figure 1: Patient flow and physician activities in a single ED pod.

In the figure above, dashed lines represent patients, and the solid lines represent the physicians moving in the ED. On the left, we have the arrivals for our two agents' patients and physicians. The patient arrival rate to the ED modeled as a non-stationary Poisson process was based on the historical data, and for physicians, their shift schedule was obtained from the partner ED scheduling system. Each patient, upon arrival to the ED, is triaged by a nurse who will then assign an ESI level to the patient as observed in the ED, which represents the patient's severity. However, it should be noted that there are a few high severity cases (car crashes, ST-Elevation Myocardial Infarction, etc.) that are taken directly to the trauma bay without triaging to provide immediate medical care. The ESI level for each patient in the model was assigned based on the retrospective data collected from the partner hospital. Patients are prioritized based on their ESI level and assigned a bed in the ED where they wait for evaluation by a physician. In case no ED beds are available, the patient waits in the waiting room, where they are prioritized based on their ESI level. The waiting room, ED beds, and trauma bays are modeled as resources where the capacities and capabilities of these resources are the same as in the partner ED.

Each physician, upon their arrival at the pod, goes to the dedicated physician workstation and checks if another physician is leaving the ED. In case another physician shift ends during the start of the new physician shift, then the arriving physician takes over the responsibility of patients from the leaving physician representing handoffs as observed in the ED. However, if no physician is leaving an ED pod, then there would be no handoffs, and the arriving physician would start taking new patients. Finally, in the case when a physician leaves the ED, and a new physician is not arriving at the ED, which happens during night shifts, the leaving physician will hand off their remaining patients to the existing physician. Following patient handoffs (if applicable), the physician would spend some time in the workstation working on patient charts, reviewing results, and then starts visiting the patients in their assigned ED bed, as necessary.

A patient's first interaction in the ED with a physician is represented as *"First Evaluation,"* during which the physician spends time examining a patient. After visiting a patient for the first time, a physician always returns to the station to update charts and order tests as needed. In case an imaging test is ordered, the physician would not meet the patient until the test is completed and the results are available for reading. The imaging process in the model is divided into a series of steps which accounts for the time from when the imaging order is placed until the image is read. After the provider places the order, the patient waits in their bed waiting for a nurse to either roll them out to the radiology suite or sometimes the imaging can be conducted as POC imaging in the patient's bed. Upon completing the imaging, the patient again waits on their bed until the results are read by the provider, after which the provider comes back to visit the patient, represented as *"Subsequent Patient Evaluations,"* where few patients need as many as 4 interactions with a physician. It should be noted that physicians can place additional imaging orders after their subsequent visits with a patient, and the imaging process will still follow the logic discussed earlier. Additionally, for the *"Subsequent Patient Evaluations,"* the care for a patient is provided by the same physician unless the physician shift ends, during which the patient is handed off to another physician. Further, as represented in the workflow, after a subsequent visit, there is a 40% chance that a physician will visit another patient before returning to the station. The time a patient spends for imaging and with a physician for each evaluation is based on the ESI level of the patient. In case of lack of data, expert opinions from ED physicians were used for modeling.

Finally, whenever there are free beds in the ED, a physician, based on their workload, will sign up a patient from the waiting room and meet them in their bed. Additionally, to replicate the actual assignment process followed at Prisma health ED, physicians working in certain pods were restricted from taking high severity patients as few pods do not have the equipment required to provide care for high severity patients. The model ensures that the physicians working in the same pod simultaneously share the patient load equally where the workload is considered balanced based on the number of patients cared for by a physician and not necessarily based on the ESI level of the patient, as that is the practice followed in the ED.

The model was developed using Rockwell Automation's Arena simulation software version 16, and after model development, the next step was to validate the model to ensure that the model replicates the partner ED. We utilized the time in the ED split by ESI level as the validation metric to ensure that patient time spent in the simulation ED did not vary significantly from the actual (observed) time in the ED data for each ESI level for the same period. The model was simulated for a three-week schedule with an additional two-day warm-up period for the model to attain equilibrium. A total of 60 replications were performed, such that the margin of error on time in the ED metric was ± 5 minutes (at α=0.05). Table 4 represents validation results where the simulated data along with the standard deviations and actual (observed) data for each ESI level. It can be observed that the largest difference between the simulated and actual data was only 6% (8 mins) which could be likely due to the small sample size for ESI 1 patients (since Level 1 has the fewest arrivals of all the patient severity levels). Further, on conducting an independent t-test, there was no significant difference (p-value = 0.90) between the simulated data and actual data, supporting that the model is representative of the partner ED.

Table 4: Simulation model validation.

| Severity | Actual Time in ED (mins) | Simulated Time in ED (mins) | Percent Difference |
|:---:|:---:|:---:|:---:|
| ESI 1 | 149 | 157±1.5 | 6% |
| ESI 2 | 261 | 260±1.9 | 0% |
| ESI 3 | 228 | 231±0.8 | 1% |
| ESI 4 | 106 | 109±1.4 | 3% |
| ESI 5 | 122 | 122±2.2 | 0% |

## 5   RESULTS

To comprehend the impact of patient delays on the patient time in the ED, we perform sensitivity analysis on two wait times (order to begin and end to read). Since there could be various factors causing these delays, including radiology department staffing shortages, lack of machines, competing demands from other departments on radiology, etc., rather than identifying the root cause leading to delays, we aimed to test multiple scenarios by reducing the wait times from 10% to 100%. First, we vary the order to begin time, and as we discussed earlier, this represents the patient waiting time from the moment a physician places the order until the imaging process is started. This delay serves as a surrogate for the availability/number of radiology resources without modeling the specific number of radiologists, machines, etc. The impact of this varying delay on patient time in the ED is represented in Table 5 below.

Table 5: Impact of imaging delays on patient time in ED.

| | Order to Begin | | End to Read | |
|---|---|---|---|---|
| **Estimated Percent Reduction in Delays** | **Percent Reduction in Patient Time in ED** | **Reduction (mins)** | **Percent Reduction in Patient Time in ED** | **Reduction (mins)** |
| 10% | 3% | 5 | 1% | 2 |
| 20% | 5% | 9* | 2% | 4 |
| 30% | 8% | 14* | 3% | 6 |
| 50% | 13% | 23* | 5% | 9* |
| 100% | 26% | 46* | 10% | 18* |

Order to begin time contributes to 50% of the total imaging process delays, and we observed that reducing this delay by 20% resulted in a 5% reduction in the patient time in the ED (9 mins). Further, a 30% and 50% reduction in order to begin time resulted in an 8% and 13% reduction in the patient time in the ED (14 mins, 23 mins). Finally, a 100% reduction in order to begin time resulted in a 26% reduction in the simulated patient time in the ED (46 mins). Although a 100% reduction is unrealistic, statistical tests revealed that even a 20% reduction could significantly (p-value < 0.05) reduce the patient time in the ED. In contrast to order to begin, the delay due to end-to-read is predominantly dependent on the availability of providers as this delay is for the provider to complete the reading. A sensitivity analysis was performed similar to the prior scenario, and the reduction in the patient time in the ED is represented in Table 4 above. The results show that reducing the percentage of end to read time did not impact patient time in the ED as strongly as a reduction in order to begin time since the end to read time contributes only to 30% of total patient delays (as opposed to 50% for the order to begin). We observed that reducing the end to read time only yields a 5% (9 mins) reduction in the patient time in the ED. Thus, minimizing order to begin should be prioritized, as shortening order to begin has a more substantial impact on patient care time. Further, on performing statistical analysis, we observed that unless a 50% reduction in the end to read time is achieved, there was no significant (p-value < 0.05) reduction in the patient time in the ED. Finally, we performed another sensitivity analysis where both order to begin and end to read delay was reduced systematically, as seen in Table 6 below.

Table 6: Impact of order to begin and end to read time on patient time in ED.

| **Percent Reduction** | **Percent Reduction in Patient Time in ED** | **Reduction (mins)** |
|---|---|---|
| 10% | 4% | 7 |
| 30% | 10% | 18 |
| 50% | 19% | 33 |
| 100% | 35% | 61 |

Based on the results represented in Table 6, reducing order to begin and end to read delay does not have a compounding effect on the overall reduction in the patient time in the ED. In fact, the average reduction in the patient time in ED when the intervals are concurrently reduced is equal to or lesser than the sum of the individual reductions for each interval. For example, when the order to begin was reduced by 30%, the average patient time in ED decreased by 14 minutes, and when the end to read time was reduced by 30%, the patient time in ED decreased by 6 minutes. However, when both delays were reduced by 30%, the total reduction in the patient time in the ED was just 18 minutes. Although reducing both delays have a higher impact than the individual reduction, the effort (cost, training, etc.) of reducing each delay should be considered to see if the benefits outweigh the efforts as the root cause for both delays could be different.

Next, we investigated the impact of bundling the image orders as opposed to placing them as separate orders for each ESI level. Since very few ESI 4 and 5 required imaging and most of those orders were single orders, they were not considered in this analysis. The rationale for conducting this sensitivity analysis was based on expert feedback and literature, which suggest that sometimes providers tend to place separate sequential orders rather than bundling these together to avoid unnecessary images and to exclude some diagnoses (Kanzaria et al. 2015). However, based on feedback from providers, there are cases where separate orders are placed when those orders could have been bundled, resulting in potential time savings that are not being realized. From our data analysis, we observed that each additional imaging order, on average, adds over 50 mins to the total time a patient spends in the ED. Table 7 below represents a sensitivity analysis performed to understand the impact of bundling image orders on average patient time in the ED.

Table 7: Impact of bundling image orders on patient time in ED.

| Scenario | Image Orders | ESI 1 | ESI 2 | ESI3 | Percent Reduction in Time in ED |
|---|---|---|---|---|---|
| Baseline | 1 | 63% | 52% | 52% | -- |
|  | 2 | 22% | 13% | 10% | -- |
|  | 3 | 5% | 3% | 2% | -- |
| S1 | 1 | 73% | 52% | 52% | -0.5% |
|  | 2 | 12% | 13% | 10% |  |
|  | 3 | 5% | 3% | 2% |  |
| S2 | 1 | 63% | 62% | 52% | -1.4% |
|  | 2 | 22% | 3% | 10% |  |
|  | 3 | 5% | 3% | 2% |  |
| S3 | 1 | 63% | 52% | 62% | -3.8%* |
|  | 2 | 22% | 13% | 0% |  |
|  | 3 | 5% | 3% | 2% |  |
| S4 | 1 | 73% | 62% | 52% | -1.9% |
|  | 2 | 12% | 3% | 10% |  |
|  | 3 | 5% | 3% | 2% |  |
| S5 | 1 | 63% | 62% | 62% | 6.6%* |
|  | 2 | 22% | 3% | 0% |  |
|  | 3 | 5% | 3% | 2% |  |
| S6 | 1 | 73% | 52% | 62% | 6.2%* |
|  | 2 | 12% | 13% | 0% |  |
|  | 3 | 5% | 3% | 2% |  |
| S7 | 1 | 63% | 52% | 52% | -2.1% |
|  | 2 | 27% | 16% | 12% |  |
|  | 3 | 0% | 0% | 0% |  |
| S8 | 1 | 90% | 68% | 64% | -9.5%* |
|  | 2 | 0% | 0% | 0% |  |
|  | 3 | 0% | 0% | 0% |  |

We first represent the baseline scenario which shows the imaging orders for each ESI level. We tested eight scenarios where S1–S3 assessed a 10% shift of imaging orders from two orders to one order for each individual ESI. In the following three scenarios, S4–S6 represent a similar 10% shift of imaging orders from two orders to one, but here the 10% shift is applied for two ESIs simultaneously. Finally, for the seventh scenario, we considered eliminating any occurrence of 3 separate image orders, and for the eighth scenario (S8), we eliminated both 2 and 3 image orders. Among the eight scenarios under consideration, only 4 had statistically significant differences (p-value < 0.05) when compared to the average time in ED for the baseline scenario. For these four scenarios, we observed that bundling the image orders helped in reducing the patient time in the ED. It can be noticed that all these significant cases involved bundling the order for ESI 3, suggesting that this is the most important type of patient class to impact the patient time in the ED. This observation makes sense because about 50% of the patient arrivals to our partner ED are classified as ESI 3. This indicates that intervention focused on ESI 3 should be prioritized, and if providers can bundle even 10% of the imaging orders placed for these patients, this could significantly reduce the patient time in the ED. Although bundling all the imaging orders is unrealistic, these analyses provide insights into the maximum impact of bundling on patient time in the ED.

## 6 CONCLUSIONS

Emergency Department crowding is a public health issue as well as a patient safety issue impacted by various factors within and beyond the control of the ED. Prior studies have identified that the imaging process delays patient care in the ED and increases the time a patient spends in the ED. Although studies have explored comparing the impact of POC testing vs. radiology suite testing, identifying root causes of imaging delays, and the number of imaging resources required to improve accessibility, none of the studies to our knowledge have explored what impact do these imaging delays have on patient time in the ED. Further, no other studies have explored the impact of bundling the image orders on patient flow in the ED.

On comparing multiple scenarios to identify the impact of two delays associated with the imaging process on the patient time in the ED, we observed that reducing the time a patient waits on their ED bed after the physician places an imaging order (order to begin) by 20% can have a significant (p-value = 0.03) reduction in the patient time in the ED. We believe that this goal could be potentially achieved by increasing staffing availability or having access to more medical devices. However, to pinpoint the exact pain points, further modeling of the Radiology department is required to identify their bottlenecks. On the other hand, investigating the time a patient waits for a provider to interpret the imaging results (end to read), we observed that to have a significant reduction in patient time in the ED, the end to read time should be reduced by at least 50%. This is primarily because the order to begin contributes more to the delay compared to the end to read. Although achieving a 50% reduction is not trivial, we believe that considering imaging requirements as a part of the staffing algorithm and a few workflow changes to notify providers where tests are ready, etc., can help reduce the end to read delay. Finally, on testing over eight scenarios for identifying the impact of bundling the image orders on patient time in the ED, we observed that even bundling 10% of imaging orders for ESI 3 patients can significantly (p-value < 0.05) reduce the patient time in the ED. Although bundling 10% of image orders reduces time in the ED for other ESI, including 1 and 2, these reductions were statistically insignificant. Further, we observed that eliminating all separate imaging orders can reduce the time in the ED by as much as 9.5%. This finding can have a significant impact on how providers conduct their practice because, as mentioned by ED providers, sometimes the orders are placed separately because they are oblivious of the delays associated with imaging.

One of the limitations of the current model is that we cannot pinpoint the exact interventions that should be taken to reduce the delays associated with imaging. However, given that order to begin delay has a significant impact on the patient time in the ED, we suspect that corrective actions should be focused on improving the staffing/resource availability in the Radiology department and have support staff roll out the patients from ED to Radiology. In future work, we plan to model the radiology department, including all its resources and staffing levels, to capture a better picture.

## ACKNOWLEDGMENTS

This study was supported by Prisma Health grant and Harriet and Jerry Dempsey Professorship in Industrial Engineering at Clemson University. We would like to thank Creative Inquiry undergraduate students from Clemson University for their support.

## AUTHOR BIOGRAPHIES

**VISHNUNARAYAN GIRISHAN PRABHU** is an Assistant Professor in the Systems Engineering and Engineering Management at the University of North Carolina at Charlotte. His research interests include developing mathematical models, simulation models and machine learning models to improve healthcare. His email address is vgirisha@uncc.edu.

**KEVIN TAAFFE** is the Harriet and Jerry Professor and Chair of the Dept. of Industrial Engineering at Clemson University. His teaching and research interests include the application of simulation and optimization in healthcare, production, and transportation logistics. Dr. Taaffe focuses on healthcare logistics problems that range from patient flow to operating room scheduling. He enjoys working on projects that bridge the gap between theoretical research and application. His email address is taaffe@clemson.edu.

**MARISA SHEHAN** is a master's student in the Dept. of Industrial Engineering at Clemson University. Her research interests include using data-driven modeling to improve healthcare. Her email address is mshehan@clemson.edu.

**RONALD G PIRRALLO** is an emergency physician and the Vice-Chair for Academic Affairs for the Upstate Prisma Health Department of Emergency Medicine. He also holds an appointment as a Professor of Emergency Medicine at the University of South Caroline School of Medicine – Greenville. His email address is Ronald.Pirrallo@Prismahealth.org.

**WILLIAM JACKSON** is an emergency physician and the Vice-Chair of Clinical Operations and Clinical Affairs for the Upstate Prisma Health Department of Emergency Medicine where he holds an appointment as an Assistant Professor of Emergency Medicine at the University of South Carolina, School of Medicine Greenville. His experience includes managing clinical operations, disaster preparedness, healthcare systems quality programming, managing large group practice risk, creating, and implementing ED design, and developing innovations in patient care delivery. His email address is William.Jackson@Prismahealth.org.

**MICHAEL RAMSAY** is an emergency medicine physician and the Director for Informatics in the Department of Emergency Medicine at Prisma Health Greenville Memorial Hospital. His email address is Michael.Ramsay@Prismahealth.org.

**JESSICA HOBBS** is an emergency medicine physician and the Medical Director for the Department of Emergency Medicine at Prisma Health Greenville Memorial Hospital. Her email address is Jessica.Hobbs@Prismahealth.org.